# Black Phosphorus Mid-Infrared Photodetectors with High Gain

Qiushi Guo[1], Andreas Pospischil[2], Maruf Bhuiyan[1], Hao Jiang[3], He Tian[4], Damon Farmer[5], Bingchen Deng[1], Cheng Li[1], Shu-Jen Han[5], Han Wang[4], Qiangfei Xia[3], Tso-Ping Ma[1], Thomas Mueller[2], and Fengnian Xia[1*]

[1]Department of Electrical Engineering, Yale University, New Haven, Connecticut 06511, USA
[2]Vienna University of Technology, Institute of Photonics, Gußhausstraße 27-29, 1040 Vienna, Austria
[3]Department of Electrical & Computer Engineering, University of Massachusetts, Amherst, Massachusetts 01003, USA
[4]Ming Hsieh Department of Electrical Engineering, University of Southern California, Los Angeles, CA 90089, USA
[5]IBM Thomas J. Watson Research Center, Yorktown Heights, NY 10598, USA

*To whom correspondence should be addressed: **fengnian.xia@yale.edu**

**Recently, black phosphorus (BP) has joined the two-dimensional material family as a promising candidate for photonic applications, due to its moderate bandgap, high carrier mobility[1-7], and compatibility with a diverse range of substrates. Photodetectors are probably the most explored BP photonic devices[8-12], however, their unique potential compared with other layered materials in the mid-infrared wavelength range has not been revealed. Here, we demonstrate BP mid-infrared detectors at 3.39 μm with high internal gain, resulting in an external responsivity of 82 A/W. Noise measurements show that such BP photodetectors are capable of sensing low intensity mid-infrared light in the pico-watt range. Moreover, the high photoresponse remains effective at kilohertz modulation frequencies, because of the fast carrier dynamics arising from BP's moderate bandgap. The high photoresponse at mid-infrared wavelengths and the large dynamic bandwidth, together with its unique polarization dependent response induced by low crystalline symmetry, can be coalesced to promise photonic applications such as chip-scale mid-infrared sensing and imaging at low light levels.**

Chip-scale mid-infrared (mid-IR) photodetectors or imaging systems can find diverse applications in industrial and environmental monitoring[13], threat detection[14], biomedical sensing[15] and free-space communication[16, 17]. The recently rediscovered black phosphorus, which has a moderate bandgap around 0.3 eV in its bulk form, represents the latest addition to the family of layered materials, and could potentially lead to high performance optoelectronic devices in the medium wavelength IR (MWIR) regime. Due to its layered structure, BP thin films are suitable for monolithic integration with traditional electronic materials such as silicon, and can be placed on different substrates, even flexible ones[9, 18]. Such great flexibility in future system integration is difficult to realize using traditional narrow-bandgap semiconductors with bulk lattices, such as mercury cadmium telluride and indium arsenide[19, 20]. Compared with gapless graphene, the moderate bandgap of BP thin films allows for suppressed dark current, and hence, low noise photodetection. BP thin films (typically in the range of 10-20 nm for our photodetectors) also offer much stronger photon absorption than single layer 2D materials such as graphene and transition metal dichalcogenides[11]. Nevertheless, existing BP photodetectors and solar cells generally exhibit a modest responsivity, making the detection of weak optical signals difficult. Moreover, previous studies have focused on the visible and near-infrared wavelength ranges only[6, 7, 10, 20].

In this work, we demonstrate BP photodetectors operating in a broad wavelength range, spanning from 532 nm to 3.39 μm. At 3.39 μm, the BP photodetector exhibits an external responsivity of up to 82 A/W.

Due to the low dark current and the high photoconductive gain, we show experimentally that the device can detect pico-watt mid-IR light even at room temperature. The impressive performances, including high photoconductive gain, high speed and at the same time low detection limit, can be leveraged to realize chip scale mid-IR sensors and imagers operating at low light levels. In addition, we show that BP's low-crystalline symmetry leads to a photoresponse related to both the incident light polarization and the photocurrent collection direction, which may be utilized for applications that require polarization sensitive detection schemes[21].

Black phosphorus crystals consist of vertically stacked, puckered layers of phosphorus atoms. In contrast to graphene, puckered black phosphorus has lower crystalline symmetry, leading to a highly asymmetric band structure. This results in an anisotropy in both electronic and optical conductivity[22], i.e. both carrier transport and light/matter interaction properties are affected by the crystalline directions. We started from fabricating BP photodetecting devices with quadruple electrodes, which enable the collection of photo-generated carriers along either $x$-(armchair) or $y$-(zigzag) direction, as shown in Fig. 1a. We first exfoliated BP thin films onto a highly doped silicon substrate, covered with 90 nm silicon dioxide. We chose BP thin films with thicknesses ranging from 10-15 nm for device fabrication. In this thickness range the absorption of infrared light significantly exceeds 10% and at the same time the channel conductance can be modulated effectively, making the realization of low off-state current possible. Thicker BP enhances light absorption, but gate control is weak due to the screening effect[23], which leads to a high dark current[2]. The upper panel of Fig. 1a presents an atomic force microscope (AFM) image of a typical device, while the lower panel shows a line-scan along the dashed white line. The transistor channel length is 2 μm and the metal contacts are 400 nm wide. The BP crystal orientations are identified using polarization-resolved Raman spectroscopy[2, 24]. Details are discussed in the **Supplementary Information I**. The source–drain current $I_{DS}$, measured while sweeping the gate voltage $V_G$ at a fixed drain bias $V_{DS}$ of 100 mV, is shown in Fig. 1 b. An on-off current ratio of ~$10^3$ is achieved. The p-branch field-effect mobilities along the x, y directions are 194 and 122 cm$^2$ V$^{-1}$ s$^{-1}$, respectively. These values are inferred from $\mu=(dI_{DS}/dV_G)(L/WV_{DS}C_{ox})$, where $I_{DS}$ is the source-drain current, $V_G$ is the back gate bias, $V_{DS}$ is the source-drain bias, $C_{ox}$ is the gate capacitance, $W$ is the width of the transistor and $L$ is the channel length. The field-effect mobility along the $x$-direction ($\mu_x$) is about 1.6 times higher than that in the y-direction ($\mu_y$), consistent with recent reports[2, 3, 25]. The lower electron current can originate from a higher Schottky barrier for electrons[1, 26], or trapping of conductive electrons at the BP/SiO2 interface.

We next measured the dependence of the photoresponse on the incident light polarization and photocurrent collection direction by employing a scanning photocurrent imaging technique. The laser spot is positioned in the center of the device by a piezo-controlled scanning mirror. The experimental details are presented in the **Methods**. Source-drain current under illumination ($I_{light}$) and dark conditions ($I_{dark}$) were both collected and the photocurrent ($I_{PH}$) was calculated by $I_{PH} = I_{light} - I_{dark}$. Figure 2 shows the photocurrent measurement results as a function of gate voltage at a source-drain bias of 200 mV with 532 nm light excitation (incident power of 230 μW). We define XX (YY) as the scenario in which light polarization and carrier collection directions are both along the $x$-($y$-) direction. When $V_G$ is swept from -10 to 20 volts, a distinct photocurrent peak is observed. The peak magnitude in the XX case is about 3 times larger than for YY. It should be noted that for one particular transport direction, the photocurrent difference by merely varying the polarization angle is a factor of 2, as can be seen from the inset of Fig. 2 for carrier collection along the x-direction. The higher carrier mobility, and therefore shorter carrier transit time ($\tau_{tr}$) along the x-direction, contribute to another factor of 1.5-1.6 in terms of photo-carrier collection and, in turn, higher responsivity.

Two distinct mechanisms can account for the photocurrent generation in BP transistor devices in this work. The first is the conventional photovoltaic effect, in which the absorption of photons with energy higher than the bandgap generates electron-hole pairs. These electron-hole pairs are separated by the electric field and drift in opposite directions towards the metal contacts, resulting in a net increase in current ($I_{PH}$). As an important signature, the photovoltaic effect is most pronounced for minimized channel doping (intrinsic regime)[27], primarily due to reduced channel carrier density and longer photo-carrier lifetime. Previously, such photovoltaic effect was found to dominate in intrinsic high mobility graphene[28, 29].

The second mechanism is the photo-gating effect. If one type of photo-generated carriers is trapped in localized states, they can act as a local gate, effectively modulating the transistor threshold voltage $V_{th}$. The photo-gating effect manifests itself as a photocurrent directly related to the trans-conductance $g_m$, i.e. $I_{PH} = g_m \Delta V_{th}$, where $\Delta V_{th}$ is the shift in threshold voltage. Microscopically, the localized states trap one type of photo-carriers and therefore prolong the recombination lifetime ($\tau_0$) of another type[30, 31]. If the lifetime of a carrier is longer than its transit time ($\tau_{tr}$), it will make several transits through the material between the contacts, provided that the contacts are able to replenish carriers drawn off at the opposite side by the injection of an equivalent carrier, which is required by the charge neutrality condition. The free carrier continues to circulate until it is annihilated by recombination, leading to a large photoconductive gain ($\tau_0/\tau_{tr}$)[32]. The photo-gating effect is prevailing in nanostructured materials, such as colloidal quantum dots[33], nanowires[34] and layered semiconductors[35]. In general, the photoconductive gain decreases when the incident light intensity increases[36, 37], since the number of localized trap states is limited. We note that photoconductive gain can still occur without considering the localized states. However, it usually requires a high mobility channel, large source-drain bias or short channel length to ensure that the transit time is small enough compared to the photo-carrier lifetime without trapping effect.

To elucidate the dominant mechanism in our device, we present device band profile models in Fig. 3a and measured the photocurrent as a function of gate bias at different excitation intensities using 532 nm light. We plot the results in the upper panel of Fig. 3b and the corresponding transistor transfer characteristic is shown in the lower panel for reference. At low incident power, the gate-dependent photocurrent curve follows the $g_m$ curve well. The gate voltage at which the photoresponse peaks is located at around the threshold voltage ($V_{th} \sim 5$ V) rather than the minimum conductance voltage ($\sim 12$ V). This behavior suggests that the dominate mechanism at low incident power is the photo-gating effect.

In semiconductors, an inhomogeneous potential distribution can result in a smearing of the sharp band edge and the formation of trap states at the band tails[38, 39]. In this case, the potential inhomogeneity can result from a random distribution of trapped charges at the interfaces and vacancies, dislocations or grain boundaries in BP itself. Trap states with energies above the Fermi level $E_F$ are empty and are able to capture electrons (electron traps). Trap states with energies below $E_F$ are negatively charged since they are filled with electrons. As a result, they are able to capture free holes (hole traps). Figure 3a illustrates the photoconductive response in the presence of localized band-tail states. First, we consider $V_G$ in the range between -10 and 0 V (region I of Fig. 3b). In this regime, the Fermi level $E_F$ is located close to the valance band as shown in the left of Fig. 3a. The concentration of free holes in the valence band is high whereas the concentration of free electrons in the conduction band is low. In dark conditions, hole trap states close to $E_F$ are therefore mostly occupied by holes whereas electron traps are nearly left empty. Under illumination, electron-hole pairs are generated and electrons are subsequently trapped (depicted by down arrows in Fig. 3a). This prolongs the photo-generated hole lifetime and leads to photoconductive gain. However, at more negative $V_G$, the BP channel is heavily p-doped and an additional barrier for holes is created in close proximity to the drain metal contact, increasing the probability of electron-hole recombination[26]. Therefore, the long carrier transit time from source to drain leads to higher probability of photo-carrier recombination and, in turn, reduced responsivity.

At higher gate voltages, $E_F$ progressively moves to higher energy and the band is almost flat. (Fig. 3a middle image, Fig. 3b region II). At $V_G \sim 5$ V, the transistor is still in its ON-state and hole traps are still mostly occupied. The trans-conductance peaks when the hole transport barrier is minimized as shown in Fig. 3b, leading to short hole transit time[26]. The trapping of photo-generated electrons renders photo-generated holes that are long-lived enough to make many transits between the contacts before annihilation. This gives rises to photoconductive gain and maximized photocurrent. However, when $E_F$ is close to the mid-gap (Fig. 3a right image, Fig. 3b region III) free hole concentration further decreases and increasing number of hole trap states becomes available to trap holes. In this scenario, both electron and hole traps states are available to capture photo-carriers. At low light levels, a large fraction of photo-generated electrons and holes get trapped first and undergo a hopping transport. Therefore, reduced photocurrent generation is observed in Fig. 3b region III due to the limited number of free carriers. It is worth noting that the hopping transport occurs even in the dark. Below a critical free hole density, hole transport in BP is mediated by thermal activation, which is revealed by the Arrhenius plot of the channel conductance under various gate voltages **(Supplementary Information Figure S2).** At higher incident light intensity, the gradual filling of trap states results in more free photo-carriers in the channel even at transistor off state. In fact, when $V_G$ is kept at 12 V, an increase of responsivity for higher incident power (Fig 3d inset) is observed. From Fig. 3b, it was observed that the photocurrent peak slightly shifts to higher gate voltage at high incident light intensity. This can be ascribed to the negative gating effect exerted by the abundant trapped photo-generated electrons, which are negatively charged. As a result, a more positive back gate voltage is needed in order to compensate this effect.

Figure 3c shows the intensity dependence of the peak photocurrent ($V_G \sim 3$ V) measured at applied biases of $V_{DS}$=100 mV and $V_{DS}$=200 mV. The clear sublinear dependence of the photocurrent on light intensity further validates the presence of the photo-gating effect. At higher illumination intensity the number of available electron-traps is reduced, leading to reduction in overall responsivity. As electron traps are filled, the number of free electrons increases and raises the probability of electron-hole recombination, which is manifested in an absorption-rate-dependent photo-carrier lifetime, $\tau_0 (F)$. The dependence of photocurrent ($I_{PH}$) on the incident power can be well described by the Hornbeck−Haynes model [37, 40]:

$$I_{PH} = q\eta \left(\frac{\tau_0}{\tau_{tr}}\right) \frac{F}{1+F/F_0} \qquad (1)$$

where $q$ is the elementary charge, $\eta$ is the absorption of BP, $F$ is the photon absorption rate in units of s$^{-1}$, and $F_0$ is the photon absorption rate when trap saturation occurs[34, 37]. The first term on the right hand side of Eq. 1 is the usual expression for the photoconductive gain, while the second term accounts for trap saturation at high excitation intensities. Figure 3d shows the photoconductive gain derived from data in Fig. 3c according to $G=(I_{PH}/P_{abs})(h\nu/q)$, where $P_{abs}$ is the absorbed power and $h\nu$ is the photon energy. The long photo-carrier lifetime combined with the short carrier transit time, which is due to the small spacing between the electrodes (2 μm), results in a photoconductive gain as high as $G=10^4$, if 15% absorption is assumed **(see Methods)**. The solid lines in Fig. 3d are the best fits to the data obtained by Eq. 1. For $V_{DS}$=100 mV, $\tau_0/\tau_{tr}=1\times10^5$ is deduced. At this bias condition, one can estimate a carrier transit time of $\tau_{tr}$=1.3 ns according to $\tau_{tr}=L^2/\mu_h V_{DS}$, where $L$ is the spacing between the electrodes and $\mu_h$ is the hole mobility, assuming a hole mobility of 300 cm$^2$ V$^{-1}$ s$^{-1}$. The photo-carrier lifetime is then estimated to be $\tau_0 \sim 0.13$ ms. Therefore, according to the above analysis, a 3-dB bandwidth of $f_{3dB}=1/2\pi\tau_0 \sim 1.2$ kHz is derived, implying that the photoconductive gain still remains effective at relatively high modulation frequencies. Under high incident power ($P_{inc} >10$ µW), the traditional photovoltaic effect plays a bigger role in photocurrent generation process, which results in higher experimental responsivity than the fitting value.

Next, we assessed the photoresponse of the BP photodetectors at 3.39 µm wavelength. A schematic view of the BP Mid-IR photodetector and a scanning electron microscope (SEM) image of the device are shown in Figs. 4a and 4b, respectively. Here, we employed a geometry containing multiple interdigitated electrodes to optimize the photo-carrier collection efficiency[41, 42]. A BP flake of ~12 nm in thickness was utilized. The interdigitated electrodes have a width of 400 nm and a spacing of 1 µm. The carrier collection was designed to be along the *x*-direction of the BP crystal and the light polarization was set in parallel to the *x*-direction. First, the photocurrent dependence on the gate voltage was measured and found to be consistent with the results shown in Fig. 2b, as shown in **Supplementary Information Fig. S3**. The absolute responsivity is calculated by photocurrent divided by the power illuminated on the device, i.e. $R = I_{PH} / (P_{inc} A_{device} / A_{laser})$, where $R$ is the absolute responsivity, $A_{device}$ is the active device area and $A_{laser}$ is the laser spot area. Fig. 4c plots the absolute IR responsivity as a function of incident light intensity for 100 mV and 500 mV source-drain biases and at the optimal gate voltage $V_G$=3 V. The photo-gating effect is clearly manifested by the reduction of the responsivity at higher optical power. A maximum responsivity of 82 A/W is achieved at 500 mV bias and 1.6 nW incident power. Figure 4d shows the dynamic gain as a function of frequency in the range from 200 to 10 kHz, obtained by modulating the laser beam with a mechanical chopper. A lock-in amplifier was used to collect the photocurrent. The 3-dB cut-off frequency decreases from 2.2 kHz to 1.1 kHz for incident optical power from 5.6 $\mu$W to 1.6 nW, as a result of free carrier reduction in the channel. The broad dynamic bandwidth renders our detector suitable for low power mid-IR imaging applications with short image acquisition time. At high chopping frequencies (>10 kHz), the detector exhibits a responsivity of ~60 mA/W regardless of the optical powers due to the photovoltaic effect, which typically offers a GHz cutoff frequency[9]. Contrary to $MoS_2$ and other semiconductors with a large bandgap and deep level traps[35, 37], we observed no significant responsivity reduction when the beam is chopped at 200 Hz compared to DC responsivity, indicating the shallow nature of traps in BP. The Hornbeck−Haynes model, which solves the trapping and de-trapping process under the steady-state condition, predicts that the responsivity reaches a plateau if the trapped charge concentration is much higher than the photo-generated free carrier concentration in the channel (low photon injection condition)[40]. Comparing our results with this model (solid line in Fig. 4c), it becomes clear that, when the incident power is less than 7 nW, the free electrons in the channel have a negligible effect on the photocurrent generation. This argument is further supported by the measured 3-dB frequency versus the incidence power (inset of Fig. 4d), since under low excitation conditions, the dynamic bandwidth is solely determined by the effective charge trapping time.

The sensitivity of our BP detector is limited by the low frequency noise of field-effect transistors. We find that the current noise of our devices is dominated by 1/*f*-noise at low frequencies (1-100 Hz), which emanates from a large number of charges being trapped and de-trapped[43] **(Fig. S4 in Supplementary Information).** We utilized a signal analyzer to monitor the source-drain current in time domain and acquired the current noise power spectral density per Hz ($S_I$) for a fixed set of $V_{DS}$ and $V_G$ in frequency domain. At frequencies higher than 10 kHz, the measured noise level is close to the shot noise limit, since the trapping states do not respond to such high frequencies. The root-mean-square (RMS) current noise amplitude ($\delta i$) is obtained by integrating $S_I$ over the measurement bandwidth and taking the square root. Figure 4e shows the measured $\delta i$ of 1 Hz bandwidth and the calculated shot noise ($(2eI_{dark})^{1/2}$) versus the back gate voltage for $V_{DS}$=500 mV. At $V_G$ of 3 V, the point at which the maximum photoresponse was obtained, we measured $\delta i$ of 0.7 nA/√Hz (the dark current level is around 5 μA). With this dark current level, the normalized noise power spectral density[44] $n \times W \times S_I / I_{dark}^2$, which quantifies the noise amplitude per unit channel length at 1 Hz, was calculated to be $6.4 \times 10^{-7}$ μm/Hz, where $W$ is the channel width (8 μm in this work) and $n$ is the number of the pair of interdigitated electrodes ($n$=4 in this work). This result is comparable with a prior report by Na, *et. al* [45] on 1/*f*-noise measurements of BP transistors. Moreover, the measured current noise level approaches the shot noise limit[46] with increased gating strength, resulting from the reduction of the dark current[47]. For any photodetector, the detection limit is assessed by the noise-equivalent-power (*NEP*), i.e. the incident power at which the signal is equal to the RMS dark current noise density ($\delta i$) within a specified bandwidth (commonly 1 Hz). The $NEP = \delta i / R$, where $R$ is the responsivity of our detector, is found to be ~8 pW/√Hz at $V_G$=3 V, indicating that mid-IR radiation in the pico-watt range can be detected above the noise level with an integration time of 0.5 s. At $V_G$ of 5 V, the *NEP* is as small as 5.6 pW/√Hz, because the dark current reduces by 58% from 3 to 5 V while the responsivity only decreases by 36%. As a result, the detector is capable of detecting low-light level signal below 10 pW in mid-IR if the gate is biased between 3 to 6 V, as shown in Fig. 4e.

In conclusion, a sensitive MWIR photodetector based on a black phosphorus transistor has been demonstrated at 3.39 μm, capable of performing low power detection in pico-watts range. The device exhibits high photoconductive gain and dynamic bandwidth in kHz range. Moreover, BP's capability to resolve incident light polarization adds another degree of freedom to photodetection devices and potentially opens new avenues to applications that require polarization-sensitive light detection schemes [21]. While traditional detecting and imaging systems focus on the capturing of color and intensity, polarization of scattered light contains rich information about the surface roughness[48], geometry[49] or orientation[50] of imaged objects. In addition, polarization contrast techniques are useful in terms of providing additional visual information in optically scattered environments, such as target contrast enhancement in hazy/foggy conditions[51]. The unique polarization sensitivity, broad spectral response and high responsivity, together with the large bandwidth and the easiness in integration with various substrates make black phosphorus a promising alternative material in mid-infrared wavelength range, in addition to traditional narrow gap semiconductors such as mercury cadmium telluride and indium arsenide.

## Methods

**Device fabrication.** The fabrication of all devices presented in this paper started with the exfoliation of BP thin films from bulk BP crystals onto a highly doped Si substrate (p-type, 0.001-0.005 Ω·cm), covered with 90 nm thermal $SiO_2$. Suitable flakes were identified by optical microscopy and their thickness was determined by atomic force microscopy. The next step was to pattern a poly (methyl methacrylate) (PMMA) resist layer for metallization using a Vistec 100 kV electron-beam lithography system. We then evaporated 3 nm Cr/35 nm Au, followed by lift-off to form the contacts. Between measurements, the samples were stored in a light-proof nitrogen box filled with a desiccant.

**Photocurrent measurements at 532 nm.** The measurements were performed with a homemade scanning photocurrent imaging setup. A green laser beam (532 nm) was focused on the device using a 20× microscope objective with long working distance. The laser spot diameter was measured to be around 3.5 μm and covered the entire device area in Fig. 1a. The incident light polarization was controlled by a half-wave plate. The absorption of BP was estimated according to a previous experimental result, in which 40% absorption was achieved in a ~ 40 nm thick BP flake in x-direction at 532 nm (2.33 eV)[11]. We assumed a constant absorption coefficient $\alpha$ (cm$^{-1}$) for BP thicknesses above 10 nm. Absorption of a 12 nm BP flake was calculated to be ~15% using the Lambert-Beer's law: $A(\omega) = 1 - e^{-\alpha(\omega)\Delta z}$ where $A(\omega)$ is the material absorption, $\omega$ is the light frequency and $\Delta z$ represents the thickness of the flake.

**Mid-IR photocurrent measurements.** Light from a Helium-Neon laser, operating at 3.39 μm, was coupled to a Fourier Transform Infrared spectrometer (FTIR) and a ZnSe objective (15×) in the Hyperion 2000 microscope was employed to focus the laser to a spot of ~11 μm in diameter. The incident laser power, which was measured to be 50 μW at maximum directly under the objective, was adjusted using multiple neutral density filters. The incident light polarization direction was adjusted using a ZnSe holographic wire grid polarizer (Thorlabs). For small signal measurements, we used a standard lock-in amplifier technique. An integration time of 1 s was chosen, which is much longer than the longest laser light modulation time constant of 5 ms ($f$ = 200 Hz). The root mean square values of the photocurrent amplitude at the lock-in output were converted to peak-to-peak values in the manuscript.

**Low frequency noise measurements**. Noise measurements were performed at room temperature in a Lakeshore probe station with micromanipulation probes. The vacuum chamber was covered with tin foil to ensure that the devices are operating in the dark. The source-drain current was amplified with a Keithley 428 low noise current preamplifier. Noise spectra at various bias conditions were acquired with an HP 3567 signal analyzer (Agilent).

## Acknowledgements

A.P. and T.M. acknowledge financial support by the by the Austrian Science Fund FWF (START Y 539-N16) and the European Union Seventh Framework Programme (grant agreement No. 604391 Graphene Flagship).

**Figure Captions**

**Figure 1 | BP field-effect transistor with quadruple electrodes.** (a) Top: AFM image of the BP photodetector. The x- and y-directions are labeled in the image. Bottom: Height profile along the white dashed line in the top panel, showing that the thickness of the BP is 12 nm. (b) Transfer characteristics of the BP field-effect transistor. $I_{DS}$ and $V_G$ are the source-drain current and back gate bias, respectively. The left and right axes represent linear and logarithmic scales, respectively.

**Figure 2 | Polarization sensitive photon detection in BP.** Effects of polarization, transport direction and back gate voltage on the photocurrent generation/responsivity in the BP transistor. The source-drain voltage is 200 mV. The incident excitation laser power is 230 µW at 532 nm. Inset: photocurrent as a function of the incidence light polarization direction for photocarrier collection along the x-direction.

**Figure 3 | Photocurrent generation mechanism in BP field-effect transistors.** (a) Schematics of device band profiles for (I) $V_G < V_{th}$, (II) $V_G \approx V_{th}$ and (III) $V_G > V_{th}$. $E_F$: Fermi level. CB: conduction band, VB: valence band. Up and down arrows denote the carrier trapping process. Black dots and open circles represent electrons and holes, respectively. (b) Upper panel: Gate- dependent photocurrent measured under a wide range of incident optical powers. The dashed line sketches the gate-dependent channel trans-conductance ($g_m$) for reference. Lower panel: transistor transfer characteristic for $V_{DS}=200$ mV. (c) Power dependence of the photocurrent at $V_{DS}=100$ mV (red symbols) and $V_{DS}=200$ mV (blue symbols). (d) Power dependence of photoconductive gain and responsivity at $V_{DS}=100$ mV (red symbols) and $V_{DS}=200$ mV (blue symbols). Solid lines: fitting results with Hornbeck−Haynes model. The absorption ($\eta$) is assumed to be 15%. Inset: power dependent responsivity measured at $V_G=12$ V.

**Figure 4 | BP mid-IR MSM photodetectors.** (a) Illustration of the BP MSM photodetector operating at 3.39 µm. The polarization of the incident laser and its incident direction are represented by yellow and green arrows, respectively. (b) False color scanning electron micrograph of the BP MSM photodetector. The green area is the BP flake with thickness of 12 nm and the golden area represents the metal contact. Scale bar: 5 µm. The spacing between the metal fingers is 1 µm and the finger width is 400 nm. (c) Power dependence of the responsivity at $V_{DS}=100$ mV (red symbols) and $V_{DS}=500$ mV (blue symbols). Solid lines: fitting results with Hornbeck−Haynes model. (d) Photocurrent amplitude versus modulation frequency under various incident powers. Inset: variation of 3-dB frequency ($f_{3dB}$) with incident power. (e) Gate-dependence of noise current density of 1 Hz bandwidth. The dashed line sketches the shot noise limit calculated from the dark current.

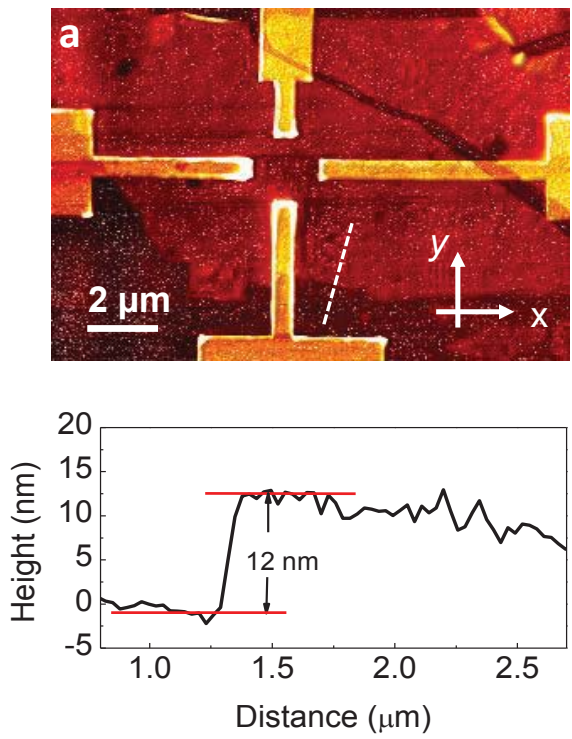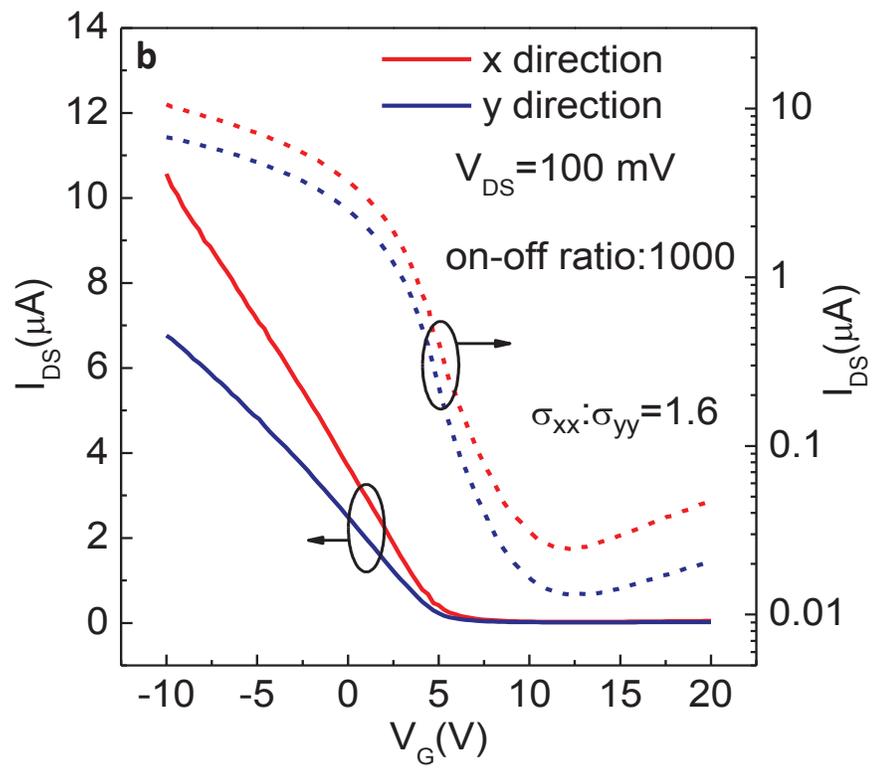

**Figure 1**

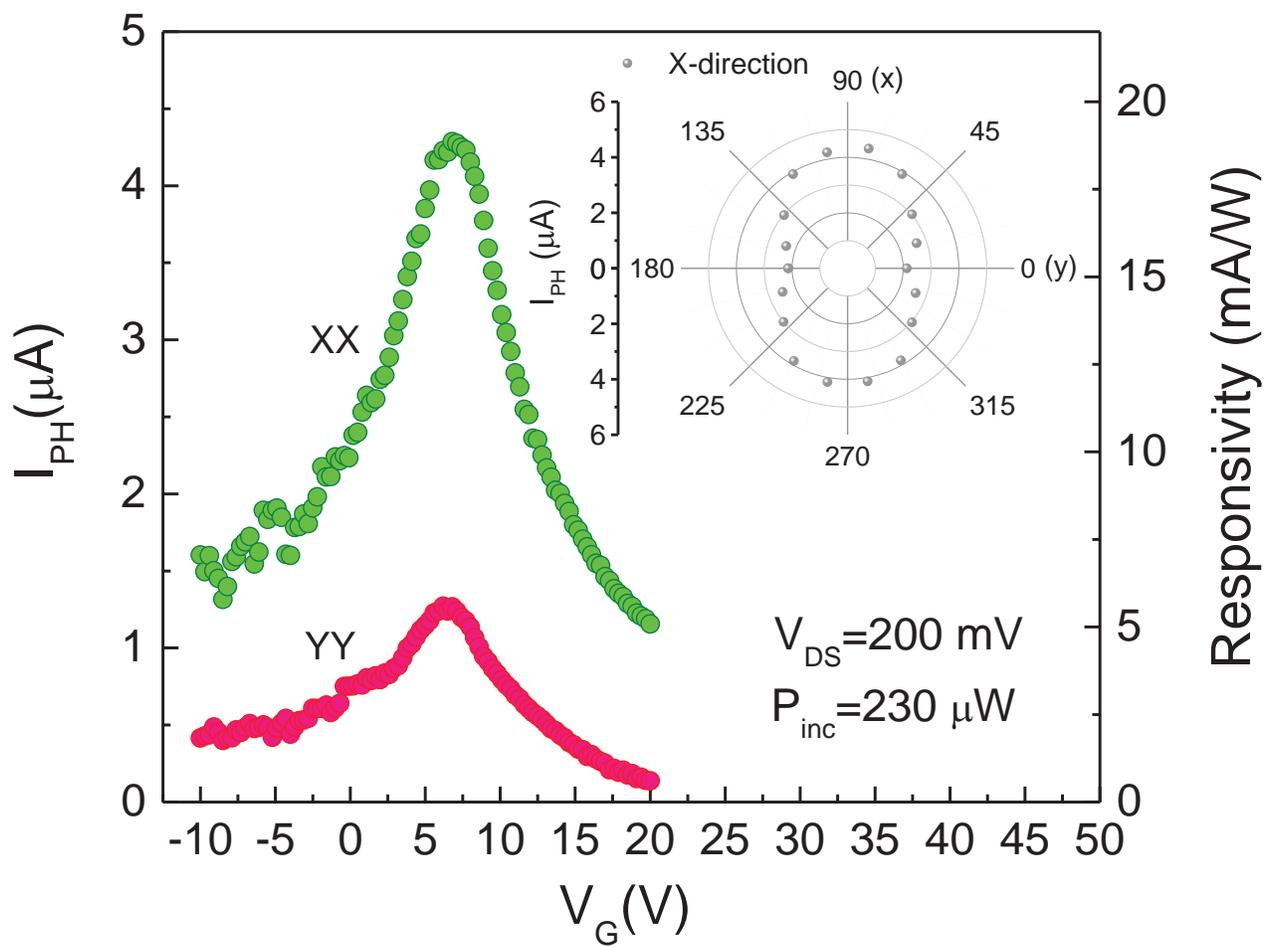

**Figure 2**

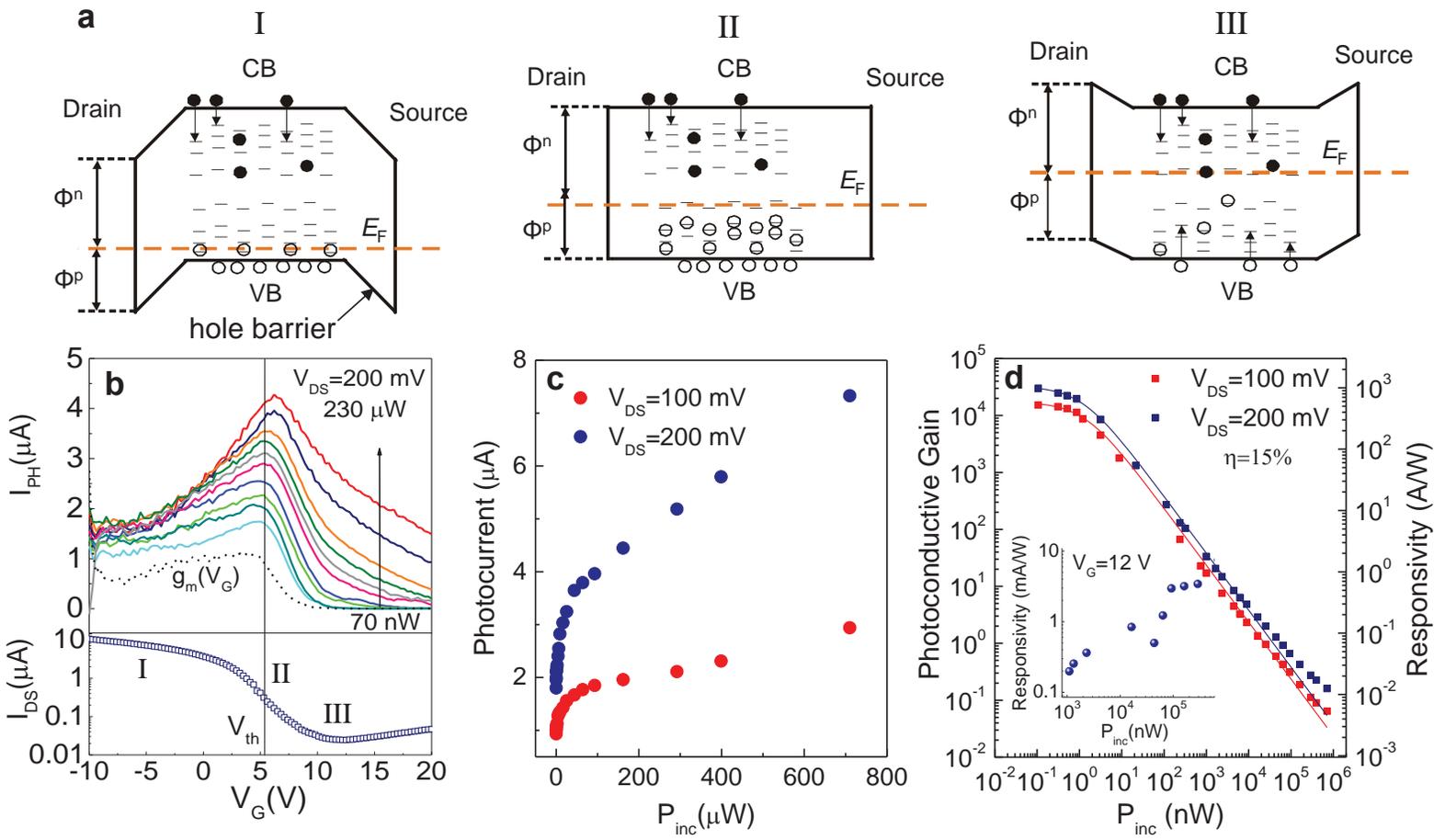

**Figure 3**

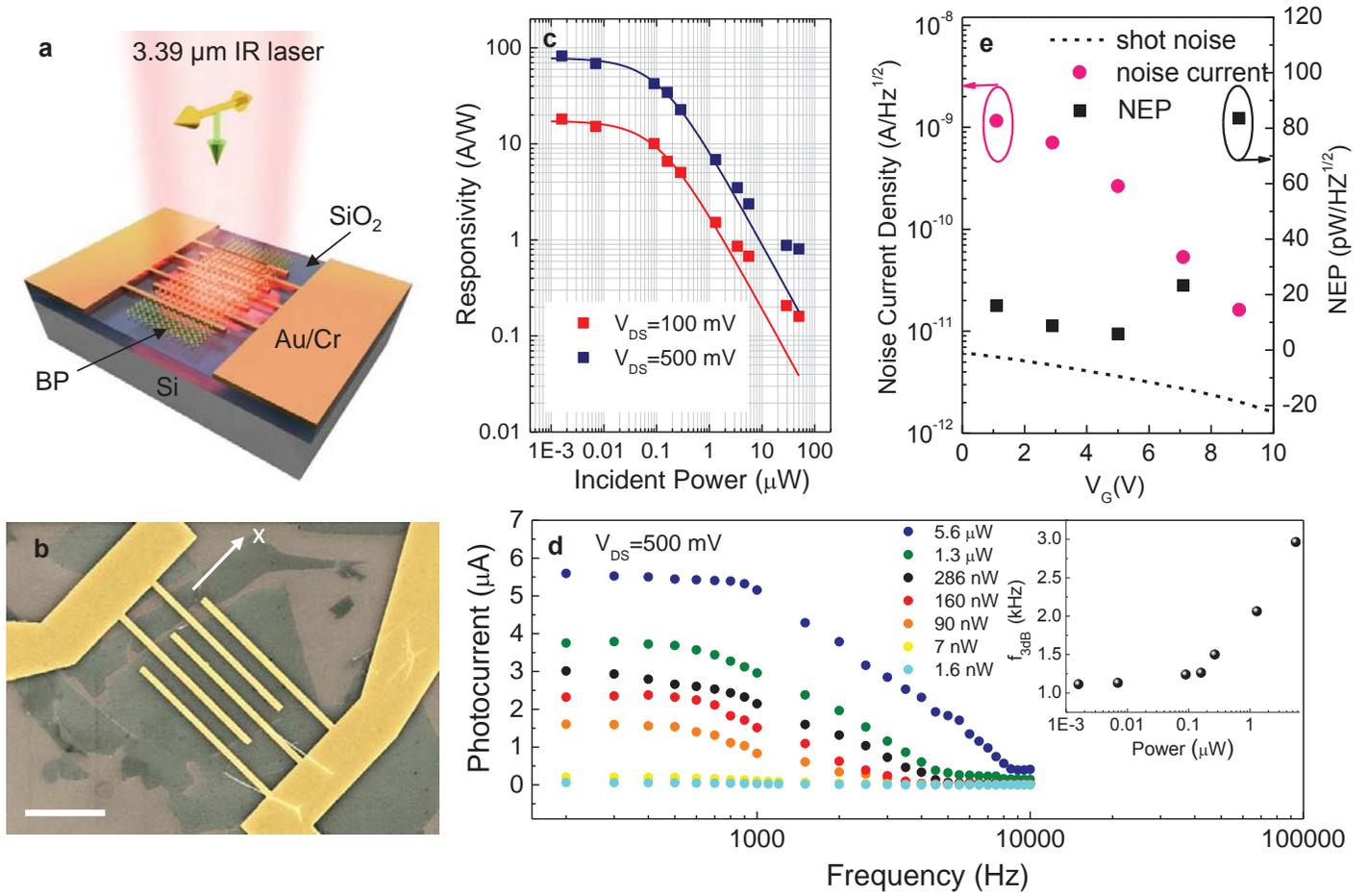

**Figure 4**